\documentclass[twocolumn,aps]{revtex4}

\usepackage{epsfig}
\usepackage{amsmath,amssymb,amsfonts}

\newcommand{\be}{\begin{equation}}
\newcommand{\ee}{\end{equation}}
\newcommand{\bea}{\begin{eqnarray}}
\newcommand{\eea}{\end{eqnarray}}

\newcommand{\bq}{\begin{eqnarray}}
\newcommand{\eq}{\end{eqnarray}}

\begin{document}
\title{Near-Milne realization of scale-invariant
fluctuations}
\author{Jo\~{a}o Magueijo$^{1,2,3}$}
\affiliation{ $^1$Perimeter Institute for Theoretical Physics, 31
Caroline St N, Waterloo N2L 2Y5, Canada\\
$^2$ Canadian Institute for Theoretical Astrophysics,
60 St George St, Toronto M5S 3H8, Canada\\
$^3$ Theoretical Physics Group, Imperial College, Prince Consort
Road, London SW7 2BZ, England
}

\begin{abstract}
A near-Milne Universe produces a very red spectrum of  vacuum
quantum fluctuations, but has the potential to produce near-scale
invariant {\it thermal} fluctuations. This happens if the energy
and entropy are mildly sub-extensive, for example if there is a
Casimir contribution. Therefore, one does not need to invoke
corrections to Einstein gravity (as in loop quantum cosmology) for
a thermal scenario to be viable. Neither do we need the energy to
scale like the area, as in scenarios where the thermal
fluctuations are subject to a phase transition in the early
Universe. Some odd features of this model are pointed out: whether
they are fatal or merely unusual should be the subject of future
investigations.
\end{abstract}


\maketitle

The near scale-invariance of the primordial density fluctuations
has become a central fact in modern cosmology and is now
well-established by observation~\cite{WMAP06cosmo}. It is still
debatable whether deviations from exact scale-invariance have been
observed (see, e.g.~\cite{raf,lid07}), and to what significance
level. But the spectral index $n_S$ cannot be substantially
different from 1, if it differs from 1 at all. The possibility
that these fluctuations might have a thermal origin has long been
entertained~\cite{peeb,steph,mag-pog}. In this paper we propose a
new thermal scenario that leads to (quasi) scale-invariance.

Thermal scenarios require new physics in order to produce near
scale-invariant fluctuations. As shown in ~\cite{param}, and
reviewed later in this paper, if one merely modifies the equation
of state of thermal matter it is found that {\it in general} the
spectral index is either $n_S=4$ (if the modes are dynamically
extracted from a sub-Hubble thermal bath) or $n_S=0$ (if there is
a phase transition, so that the final spectrum is proportional to
the equal-temperature spectrum). Therefore, prima facie, thermal
fluctuations lead to spectra that are either too blue or too red.
Independently of this result, we shall call these two types of
scenario Type A (no phase transition) and B (phase transition).

New physics, naturally, may come to the rescue, modifying the
standard calculation and rendering thermal scenarios viable.
Although this feature necessarily makes them more unconventional,
it is also precisely why they may be an interesting gateway into
phenomenological quantum gravity and string theory. For example,
in type A scenarios, loop quantum cosmology~\cite{lqc} is known to
introduce enough novelties in the gravitational sector to allow
for thermal fluctuations to produce (quasi)
scale-invariance~\cite{param}. In Type B scenarios, where there is
a phase transition ``photographing'' the equal-temperature
spectrum, one needs to deform the matter sector instead. If the
energy becomes strongly non-extensive, specifically proportional
to the area, then the equal temperature thermal spectrum is
scale-invariant. This might happen in an holographic phase of loop
quantum gravity~\cite{holo} or in the Hagedorn phase for certain
types of string gas~\cite{hag}.

In this paper we explore an intermediate possibility, combining a
type A scenario (where the modes imprinted outside the horizon
froze-in at different temperatures), with the main feature
required to render type B models viable: non-extensiveness of the
energy. This allows us to considerably soften the assumptions
about new physics. In contrast with the work of~\cite{param}, we
assume that the Einstein equations are not modified, to zeroth or
first order. In analogy with the work of~\cite{hag,holo} we let
the energy and entropy be non-extensive; but we find that they
need only be mildly sub-extensive for near scale-invariance to be
achieved, near the Milne point, $w\approx =-1/3$.

The case against thermal fluctuations at first seems utterly
damning. It is well known that thermal matter does not need to
have the usual equation of state $w=1/3$. For example deformed
dispersion relations have been discussed that induce a variety of
$w=p/\rho$, in particular $w<-1/3$. One can then have scenarios
where the modes start inside the horizon, in a thermal state, and
then are pushed out of the horizon, where they fall out of thermal
equilibrium. We can consider either contracting Universes with
$w>-1/3$, or expanding Universes with $w<-1/3$. The question is
then what is the spectrum of such fluctuations, as imprinted
outside the horizon?

The standard lore is to start by working out the fluctuations
inside the horizon, considering a generic box of volume $V$.
From the partition function
\be
Z={\sum_r} e^{-\beta E_r} \ ,
\end{equation}
(where $\beta = T^{-1}$) we find that the total energy $U$
inside volume $V$ is given by:
\begin{equation}
U={\langle E\rangle}={{\sum _r} E_r e^{-\beta E_r} \over {\sum_r}
e^{-\beta E_r}}=-{d\log Z\over d\beta}
\end{equation}
whereas its variance is given by
\begin{equation}\label{varE}
\sigma^2_E={\langle E^2\rangle}-{\langle E\rangle}^2={d^2\log
Z\over d\beta^2}= -{dU \over d\beta}=T^2c_V
\end{equation}
where $c_V$ is the specific heat at constant volume. This establishes a
very simple relation between thermal fluctuations and the specific
heat at constant volume, and this beautiful relation has been known since
the XIX century, and is textbook material~\cite{kittel}.

It is more common to discuss fluctuations in Fourier space. This
can be achieved by noting the proportionality between the variance
$\sigma^2_g(R)$ (in any quantity $g$) defined in position space
and smeared on a scale $R$, and the ``dimensionless power
spectrum'' ${\cal P}_g$. Specifically $\sigma^2_g(R)={\langle
\delta g^2\rangle}_R\approx {\cal P}_g(k_R=a/R)$. This formula has
been questioned in some regimes~\cite{joy,dav}, but we shall stay
well away from the problem cases. Then, using ({\ref{varE}) we
have \be{\cal P}_{\delta\rho}(k)\sim {\langle
\delta\rho^2\rangle}_{R=\frac{a}{k}}={1\over R^6}{\langle \delta
E^2\rangle}_R ={\sigma^2_E(R)\over R^6}={T^2\over R^6}c_V\, . \ee
We must now find out the associated metric fluctuations. The most
robust description is in terms of the (spatial) curvature
perturbation in the comoving gauge, $\zeta$. However, the
Newtonian potential in the longitudinal gauge, $\Phi$, is also
commonly in use. Outside the horizon, $\zeta$ is proportional to
$\Phi$ in expanding Universes with $-5/3<w<1$. In this regime they
both ``freeze-in'', have the same spectrum and can be used
interchangeably. Specifically $\zeta\approx 2\Phi$ near the Milne
point to be studied in this paper.

In finding the potential/curvature left outside the horizon we
must compute it from the matter fluctuations at horizon crossing.
Depending on which gauge one uses the relevant expressions look
different, but they all lead to the same result: at horizon
crossing the potential is proportional to the density contrast
$\delta=\delta\rho/\rho$. The proportionality constant differs by
a factor of order one but the spectral index is always the same.
For example in the longitudinal gauge the Hamiltonian constraint
gives: \be k^2\Phi +3H(\dot\Phi+H\Phi)=4\pi G a^2
\delta\rho^{l.g.} \ee whereas using the comoving density  we find
\be \label{poisson1} k^2\Phi= 4\pi G a^2 \delta\rho^{c.g}, \ee
(see Eqn. 14.151 of~\cite{liddle} or Eqn. 4.3 of~\cite{ks}).
Whichever of these expressions  one uses for $\Phi$ at horizon
crossing (defined as $k\sim aH$ or $k\eta\sim 1$) the result: \be
{\cal P}_\Phi\sim {a^4{\cal P}_{\delta\rho}\over k^4}\sim
{a^4\over k^4} {\left[{T^2\over R^6} c_V\right]}_{R=\frac{a}{k}}=
{k^2\over a^2}T^2 [c_V]_{R=\frac{a}{k}} \ee If we now assume the
extensive nature of the energy we have $c_V=\rho'(T)R^3$ (with
$\rho'=d\rho/dT$), so finally \be {\cal P}_\Phi\sim {a\over k}T^2
\rho'\, , \ee that is, the fluctuations at a fixed temperature are
white noise ($n_S=0$), and have an amplitude that only depends on
the Stephan-Boltzmann law, relating energy density and
temperature. We shall assume that $\rho\propto T^\zeta$, where
$\zeta$ is a parameter.

As the $\Phi$ modes leave the horizon their amplitude gets fixed
at whatever thermal amplitude they have at crossing, that is:
\be\label{xing} {\cal P}_\Phi(k)\sim {\left[{a\over k}T^2
\rho'\right]}_{k=aH}\, . \ee Using the Friedman equation
$H^2\propto\rho$ we therefore have \be\label{PT} {\cal
P}_\Phi(k)\sim {\left[T^2\rho'\over \sqrt{\rho}\right]}_{k=aH} \ee
where $k=aH$ specifies a relation between a given comoving $k$
leaving the horizon at a given time, and the temperature, thereby
allowing the inversion of the right hand side as a function of
$k$. Eqn.~(\ref{PT}) implies: \be\label{nPofT} {d\ln {\cal
P}_\Phi\over d\ln T}=1+{\zeta\over 2} \ee The relation between
$k=aH$ and the temperature, however, depends on both the equation
of state $p=w\rho$ and $\rho\propto T^\zeta$. Using the Friedmann
equation we have $k=aH\propto a\sqrt{\rho}$, and  since
$\rho\propto 1/a^{3(1+w)}$, we may derive \be\label{aofT} a\propto
T^{-\zeta\over 3(1+w)}. \ee Therefore: \be\label{nkofT} {d\ln
k\over d\ln T}= {-\zeta\over 3(1+w)}+{\zeta\over 2}
={\zeta(1+3w)\over 6(1+w)} \ee and the spectral index is
%, but always noting that we are
%assuming that the modes leave the horizon, so that $w<-1/3$.
\be\label{n-1} n_S-1={d\ln {\cal P}_\Phi\over d\ln k}= {d\ln {\cal
P}_\Phi\over d\ln T}{d\ln T\over d\ln k} =3{2+\zeta\over
\zeta}{1+w\over 1+3w} \ee A similar result can be obtained by
considering the outside and inside the horizon expressions for
$\Phi$ and matching them at $k\eta\sim 1$ (as in the heuristic
argument presented at the start of~\cite{andrew}).

Immediately two interesting candidates for scale invariance stand
out. Firstly $\zeta=-2$, that is $\rho\propto1/T^2$; this may lead
to scale-invariance because the amplitude of the frozen-in thermal
fluctuations does not depend on the temperature when they leave
the horizon (c.f. Eqn.\ref{PT}). Secondly $w=-1$: this could lead
to scale-invariance because $\rho$ does not change, and so neither
does the temperature or amplitude of the fluctuations as they
leave the horizon. This is the scenario studied
in~\cite{rob,ncinfl1}.

However further conditions apply. It's been
noted~\cite{verl,youm,dias} that the equation of state $p=w\rho$
and the Stephan-Boltmann law $\rho=\rho(T)$ are linked by a
thermodynamical relation. The argument assumes that energy and
entropy are extensive. Consider the first law of thermodynamics:
\be\label{1stlaw} dU=-pdV+TdS \ee If the energy $U$ and entropy
$S$ are extensive, then $U(\lambda V,\lambda S)=\lambda U(V,S)$.
Taking a derivative with respect to $\lambda$ at $\lambda=1$, and
using(\ref{1stlaw}) we arrive at the Euler relation
\be\label{euler} U=-pV+TS \ee so that defining $\rho=U/V$ and
$s=S/V$ we have \be s={p+\rho\over T} \ee We can now prove that
$s=dp/dT$ in a variety of ways, e.g. introducing the free energy
$F=U-TS=F(V,T)$, so that $dF=-pdV-SdT$. This leads to the
integrability condition: \be s={\left(\partial S\over\partial
V\right)}_T={\left(\partial p\over
\partial T \right)}_V
\ee Thus the expression \be\label{prhot} {dp\over dT}={p+\rho\over
T} \ee If $w$ is a constant we obtain that $\rho\propto T^\zeta$
with \be\label{zetaw} \zeta=1+{1\over w} \ee This relation implies
that  $\zeta=0$ for $w=-1$: ``deformed'' radiation may behave like
a cosmological constant, but then the specific heat vanishes and
there are no thermal fluctuations at all. For $\zeta=-2$ it leads
to $w=-1/3$, that is, {\it exactly} a Milne Universe. This is
merely a borderline case for solving to the horizon problem: the
comoving horizon does not increase but neither does it increase.
In fact this relation implies that $n_s=4$ for all values of $w$
and clearly it must be broken if this type of scenario is to be
viable.

This can be achieved if the energy and entropy are sub-extensive
as we now prove. Let's suppose that energy and entropy they scale
like \bea
U&=&\rho_C(T)V^{1-\gamma}\\
S&=&s_C(T)V^{1-\gamma}
\eea
with $\gamma>0$. Then the Euler relation (\ref{euler}) is modified,
creating a parallel with the Casimir energy, as pointed out in~\cite{verl}.
Specifically, we can prove, along the same lines as before, that now
\be
(1-\gamma)s_C=\frac{p_C+\rho_C(1-\gamma)}{T}
\ee
with pressure defined as $p_C(T)=pV^\gamma$.
The Casimir energy defined for this model is
\be
E_C=3(U+pV-TS)=3\gamma F
\ee
where $F=U-TS$ is the free energy. As before,
using $dF=-pdV-SdT$ we can
prove that
\be
\frac{dp_C}{dT}=(1-\gamma)s_C
\ee
and this now leads to
\be
\frac{dp_C}{dT}=\frac{p_C+\rho_C(1-\gamma)}{T}
\ee
For $p_C=w_C\rho_C$ and $\rho_C\propto T^{\zeta_C}$
we therefore have
\be\label{modconst}
\zeta_C=1+\frac{1-\gamma}{w_C}
\ee
a constraint to be contrasted with (\ref{zetaw}).

This modification is sufficient to break both the white-noise nature of the
spectrum at fixed temperature, and the conclusion that $n_S=4$
if we extract modes at different temperatures. Specifically
the equal-time power spectrum is
\be
{\cal P}_\Phi=T^2\rho_C'\left(\frac{a}{k}\right)^{1-3\gamma}
\ee
and we see that for $\gamma =2/3$ the (equal time)
spectrum is scale-invariant,
as derived in the holographic scenario of~\cite{holo}.

For more modest values of $\gamma$ the spectrum is only slightly
non-white. However we can now find a condition for scale
invariance {\it outside the horizon}. If the modes freeze for
$k=aH$ and if the Einstein equations (zero and first order) are
unmodified, then outside the horizon we have \be \frac{d \ln {\cal
P}_\Phi}{d \ln T} = 1+\frac{\zeta_C}{2}(1+3\gamma) \ee so that the
condition for scale-invariance is \be
\zeta_C=\frac{-2}{1+3\gamma}\, . \ee Making use of the modified
constraint (\ref{modconst}) this translates into \be\label{result}
w_C=-\frac{(1-\gamma)(1+3\gamma)}{3(1+\gamma)} \approx
-\frac{1}{3}(1+\gamma) \ee so that for $\gamma>0$ (sub-extensive
energy) we can have $w<-1/3$. Naturally we can also fix the
parameter to make the spectrum as close or deviant from
scale-invariance as required, with result \be\label{result1}
n_S-1=3\frac{1+\gamma\frac{3(w_C-\gamma)-4}{1+3w_C}}
{1-\frac{\gamma}{1+w_C}} \ee

What might go wrong with such a scenario? Most obviously there is
the issue of the validity of the whole set up. If $w<0$, then
$\zeta<0$. If $T>0$ (and $\rho>0$) this would entail a negative
specific heat $c_V$, something that signals the break down of the
canonical ensemble and the onset of a phase transition(see,
e.g.~\cite{lynden-bell}). Indeed expression (\ref{varE}) is the
standard proof for the fact that the specific heat must be
positive~\cite{schro}. This conclusion, however, can be bypassed
if we are prepared to accept negative temperatures~\cite{phant}.
Negative temperatures have been discussed in the context of
nuclear spin systems, and the usual formulae of thermodynamics and
statistical physics can be easily adapted to them~\cite{rams}.
Much the same arguments that can be made with phantom matter,
then, apply here~\cite{phant}.

Perhaps the strongest constraint on this type of scenario may
derive from the number of e-foldings required to produce the whole
observable Universe. If $w$ is very close to $-1/3$ this can be
vastly larger than in the deSitter case, and since $H$ and $T$ are
varying obvious constraints ensue, telling us how far we must be
from  the Milne Universe, that is how large $|w+1/3|$ must be.
This translates into a constraint on the necessary $\gamma$ (see
Eqn.~(\ref{result}) and (\ref{result1})), so that the
sub-extensiveness may not be mild at all. If if this is the case,
however, we argue that these scenarios are still worth pursuing.

To summarize we have investigated a thermal scenario that leads to
near scale-invariant fluctuations. It bypasses the usual result
$n_S=4$ by allowing the energy and entropy to be mildly
sub-extensive. Corrections to Einstein's gravity do not need to be
invoked if $w$ is smaller than $-1/3$, but not by much. Neither
does the energy need to go like the area, as in other (more
speculative) work. Indeed we advocate this model as the most
conservative viable thermal scenario in existence. It is curious
that this region of parameter space, so unsuitable for  the vacuum
fluctuations (for which, then, $n_S\sim -\infty$), is so
hospitable to thermal fluctuations.

There are other ``conservative'' ways in which the standard
$n_S=4$ result may be avoided. If we accept a two fluid system,
one responsible for the fluctuations, the other for the background
evolution, then a viable scenario also follows without any extra
assumptions. However such scenarios are extremely fine tuned, as
we shall detail in a future publication. The scenario presented
here may fare much better, with only the simple extra assumption
of a Casimir contribution to the energy of the Universe.

I'd like to thank Robert Brandenberger, Parampreet Singh and Andrew
Tolley for helpful discussions.  Research at
PI is supported in part by the Government of Canada through NSERC and
by the Province of Ontario through MEDT.

\end{document}